# A simple model for the evolution of molecular codes driven by the interplay of accuracy, diversity and cost

**Tsvi Tlusty**

*Department of Physics of Complex Systems, Weizmann Institute of Science, Rehovot, Israel, 76100,*

*E-mail: tsvi.tlusty@weizmann.ac.il*

**Abstract.** Molecular codes translate information written in one type of molecules into another molecular language. We introduce a simple model that treats molecular codes as noisy information channels. An optimal code is a channel that conveys information accurately and efficiently while keeping down the impact of errors. The equipoise of the three conflicting needs, for minimal error-load, minimal cost of resources and maximal diversity of vocabulary, defines the fitness of the code. The model suggests a mechanism for the emergence of a code when evolution varies the parameters that control this equipoise and the mapping between the two molecular languages becomes non-random. This mechanism is demonstrated by a simple toy model that is formally equivalent to a mean-field Ising magnet.

## 1. Introduction

The information-processing circuitry of the cell uses molecular codes to translate symbols written in one type of molecules into another molecular language. For example the genetic code translates DNA base-triplets into amino-acids [1-3]. A molecular code is an information channel that utilizes noisy recognition interactions to map molecular symbols to their respective meanings (figure 1). Clearly, molecular codes that can reduce the impact of recognition errors while accurately and efficiently translating information can provide the organism with an evolutionary edge over its competitors. While the actual evolutionary dynamics of molecular codes may be complicated, three basic forces that could drive this evolution are apparent: Molecular recognition may misread a symbol as a similar, yet different, symbol and thereby misinterpret it as carrying an erroneous meaning. The code may reduce the deleterious impact of such a misreading error, termed error load, by assigning similar meanings to similar symbols. Thus, minimizing the error-load is an evolutionary force that tends to "smooth" the code, in the sense that similar symbols that are likely to be confused tend to bear also similar meanings [2-5]. At the extreme, this force will drive all the symbols to encode the same meaning but, of course, a one-meaning vocabulary is too limited for efficient information transfer. In the case of the genetic code, for example, a diverse amino-acid vocabulary is essential for the synthesis of functional proteins. Thus diversity is a second evolutionary force that counteracts error-load in driving the code to be as heterogeneous as possible. Finally, one must remember that the code is realized in molecules, which cost the organism materials, energy and time to synthesize and maintain. Therefore, the third evolutionary force to consider is this cost of the molecular coding apparatus.

The natural theoretical framework to treat molecular codes is that of information and coding theory. In particular, the branch of rate-distortion theory [6-8] deals exactly with the problem considered here of how to optimize a noisy information channel. Formally, the channel or the code is defined as a probabilistic mapping of input symbols to output meanings (figure 1). To optimize the code, one tries to estimate the symbol-meaning mapping that maximizes the quality of the code under the constraint of a given bound on its cost. Within the general equivalence of information theory and statistical mechanics [6-9], the quality of the code plays the role of an elastic or surface energy that tends to "smooth" the code by driving neighboring symbols to encode similar meanings. The cost is analogous to the entropy and measures the symbol-meaning correlation and thus the specificity of molecular recognition required to establish this correlation. Hence, optimizing the code is equivalent to minimizing a free energy whose physical degrees of freedom are the mapping probabilities. On that ground, the code optimization problem can be likened to standard physical systems, such as binary mixtures or mean-field Ising magnets and this analogy provides insight into the evolution of molecular codes.

Recently, we have used the rate-distortion framework to introduce a theory of molecular codes [2, 3, 10]. In short, one describes the molecular code as a two-way channel that first encodes a meaning as a symbol, then reads the symbol and finally decodes back its meaning. The adjective 'two-way' indicates the coupling between a forward mapping from meanings to symbols (the encoder) and a backward mapping from symbols to meanings (the decoder). While this general model explained the emergence of the code as a consequence of the balance between the cost and the quality of the coding system, the resulting formalism is inherently non-linear due to the encoder-decoder coupling and could be resolved only by various approximations (this coupling corresponds to code-message co-evolution, see [5] and references therein). Moreover, this non-linearity made the contributions of the evolutionary forces due to error-load and diversity practically inseparable.

This motivates us to present here a simple model that treats the molecular code as a noisy one-way channel that reads symbols and decodes their meanings. The simplicity of the model allows us to derive intuitive expressions for all three evolutionary forces due to the error-load, the diversity and the cost of the channel and the code fitness is a linear combination of these forces. It is shown that when the cost is dominant, the mapping is totally random, that is symbols and meanings are uncorrelated. Changing the equipoise of the three forces leads to the emergence of a molecular code, when the mapping between symbols and meanings becomes non-random. The basic features of the model are demonstrated by an instructive toy model of a code that maps two symbols to two meanings, which is analogous to a mean-field Ising magnet. The correlation between symbols and meanings is measured by the same order parameter that measures the average orientation of the magnet. Specifically, the emergence of the code when the correlation becomes non-zero is analogous to the familiar second-order phase transition of the magnet. We conclude by discussing the impact of the symbol-space topology and sketching a scheme for an experimental test.

## 2. The fitness of molecular codes

A molecular code can be seen as a noisy information channel (figure 1) that reads a molecular symbol as input and churns out a meaning at its output [2, 3]. The rate of information transfer through the channel is measured by the specificity of assigning meanings to symbols. If the meaning at the output is drawn randomly and is therefore uncorrelated with the input symbol, then the assignment is non-specific and the information rate vanishes. Information is conveyed only when the input and output are correlated, even partially, by specific molecular recognition. However, such specificity requires the synthesis of molecular recognizers with matching binding sites. As we discuss below, a code emerges exactly at the point when the benefit of specificity and the consequent non-zero information flow surpasses the cost of constructing the coding system.

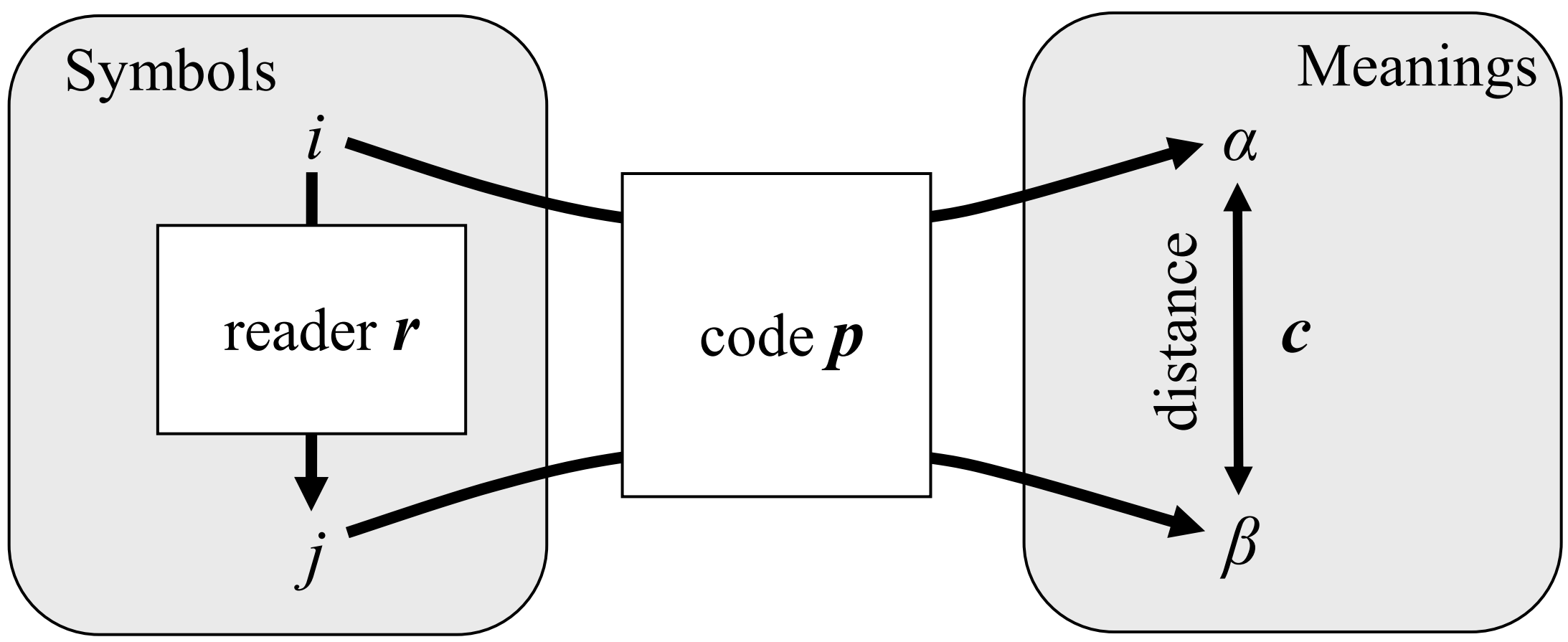

Fig. 1. **Molecular codes as noisy information channels.** The code maps the space of symbols at the input of the channel (left) into the space of meanings at its output (right). The channel consists of two stages: First a symbol $i$ is read by the error-prone molecular "reader". Sometimes, $i$ might be misread as the symbol $j$. This is characterized by the reading matrix $\boldsymbol{r}$, which lists all the probabilities for such reading/misreading events (see text). At the second stage, the read symbol $j$ is mapped to a meaning $\beta$ according to the code matrix $\boldsymbol{p}$. Misreading a symbol at the input of the channel, for example replacing $i$ by $j$ (left), may result in a wrong meaning at its output, for example receiving $\beta$ instead of $\alpha$. The corresponding fitness load is measured by the distance matrix $\boldsymbol{c}$ (right).

To quantify the evolution of a molecular code, let us consider a noisy information channel, which first reads symbols from the space of $n_s$ symbols, $i, j, k$… and then maps the read symbols into the space of $n_m$ meanings, $\alpha, \beta, \gamma$… (figure 1). The channel is a stochastic map specified by the *code matrix* $\boldsymbol{p}$ of the $n_s \times n_m$ probabilities $p_{i\alpha}$ that a symbol $i$ encodes a meaning $\alpha$. Each row of the code matrix satisfies probability conservation, $\Sigma_\alpha\, p_{i\alpha} = 1$. When the cost of specificity is dominant, the relation of symbols and meanings is non-specific. At this non-coding state, any symbol is equally likely to encode any of the available meanings and the corresponding code-matrix is therefore uniform, $p_{i\alpha} = 1/n_m$. A code emerges when the code matrix becomes non-uniform, signifying that at least some of the symbols tend now to encode preferred meanings [2, 3, 11, 12]. Returning to the physical analogy, each of the $n_s$ symbols corresponds to a physical degree of freedom, a "spin", which may take one out of $n_m$ possible "states" given by the meanings. Within this analogy, the code matrix $p_{i\alpha}$ is simply the ensemble-average probability that the $i$-th "spin" is in the $\alpha$-th state. Hence, a two-state system is equivalent to the Ising magnet, whereas the general $n_m > 2$ case is analogous to the Potts model [13]. We note that the present one-way formulation of the coding system disregards some important aspects of symbol-meaning co-evolution [5] due to encoder-decoder coupling in the two-way formulation [2, 3]. Specifically, the flexibility of reassigning meanings to symbols – which is important at later stages of the code's evolution, far from the coding transition – is reduced.

Next, we derive simple expressions for the three evolutionary forces as a function of the code-matrix $\boldsymbol{p}$ and begin by estimating the error-load. The error-prone molecular reading apparatus may misread a symbol $i$ as $j$ (figure 1). The molecular "reader" is characterized by the $n_s \times n_s$ reading matrix $\boldsymbol{r}$ of the probabilities $r_{ij}$ for all such misreading events. The diagonal of this matrix, $r_{ii} = 1 - \sum_{j \neq i} r_{ij}$, lists the probabilities to read the symbols correctly. The symbol space with its reading matrix exhibit the geometry of a graph, in which the nodes are the symbols and an edge connect two symbols if they are likely to be confused. Misreading a symbol may result in a wrong meaning and the impact of such errors is measured by the $n_m \times n_m$ distance matrix $c_{\alpha\beta}$ (whose diagonal elements vanish $c_{\alpha\alpha} = 0$). The error-load due to an $i \rightarrow j$ misreading is the average distance between all possible pairs of meanings encoded by the symbols $i$ and $j$, $\sum_{\alpha,\beta} p_{i\alpha}\, p_{j\beta}\, c_{\alpha\beta}$ [2, 3, 14]. The overall *error-load* $L$ is simply the sum of the error-loads between all pairs of adjacent symbols, weighed by the probability that such a misreading occurs, $r_{ij}$,

(1) $$L = \sum_{i,j,\alpha,\beta} r_{ij}\, p_{i\alpha} p_{j\beta} c_{\alpha\beta}.$$

The error-load (Eq. 1) resembles the spin-spin interaction energy of a ferromagnet, where the reading matrix $\boldsymbol{r}$ indicates the range of the "ferromagnetic" interaction and $\boldsymbol{c}$ is its magnitude. The error load $L$ is maximal if two interacting spins, say $i$ and $j$, encode the most distant meanings and vanishes if they encode exactly the same meaning $\alpha$, $p_{i\alpha} = p_{j\alpha} = 1$. Apart from its smoothing effect, the error-load $L$ also drives each of the symbols to prefer one particular meaning over a mixture of several meanings. This is evident from the contributions to the error-load due to correct reading events $i \rightarrow i$, $r_{ii} \sum_{\alpha,\beta} p_{i\alpha}\, p_{i\beta}\, c_{\alpha\beta}$, which vanish if and only if the symbol $i$ is mapped into a single meaning. These contributions account for the benefit of having a mapping that can accurately represent specific meanings.

The tendency of the error-load to homogenize the code-matrix is counteracted by the need for *diversity*. One can estimate the fitness benefit of diversity by the average distance $D$ between the meanings encoded by all pairs of different symbols,

(2) $$D = \sum_{i,j,\alpha,\beta} \left(1-\delta_{ij}\right) p_{i\alpha} p_{j\beta} c_{\alpha\beta}.$$

The diversity $D$ vanishes when all the symbols encode the same meaning and is maximal if the symbols encode the most distant meanings. The diversity is "anti-ferromagnetic" in the sense that it drives the spins to different as possible states. It is evident from Eqs. 1-2 that while the error-load $L$ "ferromagnet" depends on the geometry of the symbol-space through the reading matrix $\boldsymbol{r}$, the "anti-ferromagnetic" diversity $D$ is geometry-independent, that is the all spins interact with each other (the $\delta$ in Eq. 2 excludes self-interaction). In the non-coding state, $p_{i\alpha} = 1/n_m$, both the diversity and the error-load are proportional to the average distance in the meaning space, $D_{nc} \sim L_{nc} \sim n_m^{-2} \sum_{\alpha,\beta} c_{\alpha\beta}$.

The specific forms of the error-load (Eq. 1) and diversity (Eq .2) forces are quite arbitrary and were chosen mainly for the sake of simplicity. In particular, the diversity term drives the decoded meanings to be as different as possible, which reflects some general notion of diversity in an abstract biochemical meaning space. Nevertheless, this disregards the possibility that there exists a biologically relevant range for such diversity and that the organism might not benefit from exceeding this diversity range. Furthermore, a too diverse vocabulary of meanings might even interfere with other considerations, such as the load of maintaining the biochemical circuitry that processes and synthesizes molecular meanings and the load of acquiring the diverse biochemical building blocks. In addition, exceeding some biocompatibility limits might make the meanings lethal to the cell. We expect such considerations to be especially relevant at later stages in the evolution of a molecular code, where the number of decoded meanings further increases [15, 16].

Molecular codes rely on recognition and binding interactions to relate symbols and meanings. In this regard, each code-matrix entry $p_{i\alpha}$ is the probability that the molecule carrying the meaning $\alpha$ binds the molecular symbol $i$. High specificity reduces the chance of misreading by increasing the chance that the molecular recognizers will bind to their designated targets and not to one of the many lookalikes. However, highly specific binding of $\alpha$ to $i$ requires higher binding energy $E_{i\alpha}$, which in general necessitates larger binding-sites [17-20]. It is plausible to assume that the cost of the code $C$ is proportional to the average size of the binding sites and therefore to the average binding energy [3]. To estimate $C$ one notes that, according to the Boltzmann distribution, the binding energy scales like the logarithm of the binding probability, $E_{i\alpha} \sim \ln p_{i\alpha}$. It follows that the average size of the binding site, and therefore the cost $C$ of the molecular code, are proportional to the average logarithm of the code matrix,

(3) $$C = \sum_{i,\alpha} p_{i\alpha} \ln\left(\frac{p_{i\alpha}}{p_{\alpha}}\right).$$

In Eq. 3, $p_{i\alpha}$ is normalized by the marginal probability $p_\alpha$ that a meaning $\alpha$ is encoded by any of the symbols, $p_\alpha = n_s^{-1}\sum_j p_{j\alpha}$. This ensures that the cost at the non-coding state vanishes $C_{nc} = 0$, because the binding is completely nonspecific and the code-matrix is uniform $p_{i\alpha} = p_\alpha = 1/n_m$. The cost $C$ measures the mutual information between the symbols and the meanings, which is the entropy reduction due to the symbol-meaning correlation in the code matrix [7]. As specificity increases the cost increases towards its upper bound, $C_{max} = n_s \ln n_s$, which occurs when every symbol encodes one meaning.

To optimize the molecular code, the three evolutionary forces, the error-load $L$, the diversity $D$, and the cost $C$, must be balanced. For the sake of simplicity, we express this balance as the maximization of the code fitness $H$, which we define as a linear combination of the three forces,

(4) $$H = -L + w_D \cdot D - w_C \cdot C.$$

The weight $w_D$ measures the relative significance of diversity with respect to the error-load, that is anti-ferromagnetic/ferromagnetic relative interaction. The cost weight $w_C$ measures the evolutionary price of the resources required by the code's machinery. In Eq. 4, the cost $C$ and the error-load $L$ are prefixed with a minus sign since they reduce the code fitness, while the diversity $D$ increases the fitness.

Since the cost is the entropy loss due to coding, the conjugate parameter $w_C$ is equivalent to a temperature. High $w_C$ "temperatures" indicate a rising price of binding sites, which drives the code-matrix to homogeneity by reducing the specificity of the underlying recognition interactions. In contrast, increasing the diversity weight $w_D$ is expected to drive the code towards higher diversity and specificity. We note that the generic linear expression for of the fitness $H$ with its Lagrange multipliers, $w_C$ and $w_D$, was chosen mainly for the sake of simplicity since there are no sound theoretical grounds for a more sophisticated functional. The optimal code $\boldsymbol{p}^*$, the one which maximizes the fitness $H$, is a function of four determinants: the reading matrix $\boldsymbol{r}$, the distance $\boldsymbol{c}$, and the two weights, $w_C$ and $w_D$. Below we discuss, within the present simplified model, how a molecular code evolves as these parameters are varied. Specifically, the emergence of a molecular code occurs at a coding transition in the noisy information channel [2, 3, 11, 12].

### 3. The evolution and emergence of a molecular code

The optimal code maximizes the overall fitness $H$ (Eq.4). To calculate the corresponding code-matrix $\boldsymbol{p}^*$, one augments $H$ with Lagrange multipliers to ensure that the $n_s$ probability conservation relations, $\Sigma_\alpha p_{i\alpha} = 1$, are satisfied [6, 7]. The resulting functional is $H_T = H + \Sigma_i \mu_i \Sigma_\alpha p_{i\alpha}$ and the optimal code-matrix $\boldsymbol{p}^*$ is located at the extremum, $\partial H_T/\partial p_{i\alpha} = 0$,

$$(5) \qquad p^*_{i\alpha} = \frac{p^*_\alpha \exp\left(-G_{i\alpha} / w_C\right)}{\sum_\beta p^*_\beta \exp\left(-G_{i\beta} / w_C\right)} .$$

In this Boltzmann-like distribution of $\boldsymbol{p}^*$, the cost parameter $w_C$ plays the role of the temperature and the effective energies are $G_{i\alpha} = 2\Sigma_{j,\beta}\, (r_{ij} - w_D(1 - \delta_{ij}))p_{j\beta}c_{\alpha\beta}$. Since both sides of Eq. 5 depend on the code matrix $\boldsymbol{p}^*$, it defines a self-consistency relation for $\boldsymbol{p}^*$, which in general requires a solution by numerical iterations [7]. At high cost "temperature", $w_C \rightarrow \infty$, the optimal code given by Eq. 5 is completely non-specific, $p_{i\alpha} = p_\alpha$. As we demonstrate below, when the cost parameter $w_C$ is decreased, the code may remain non-specific for some range of $w_C$. Then, below a certain critical value $w_C$*, the code matrix undergoes a sharp coding transition when it becomes specific.

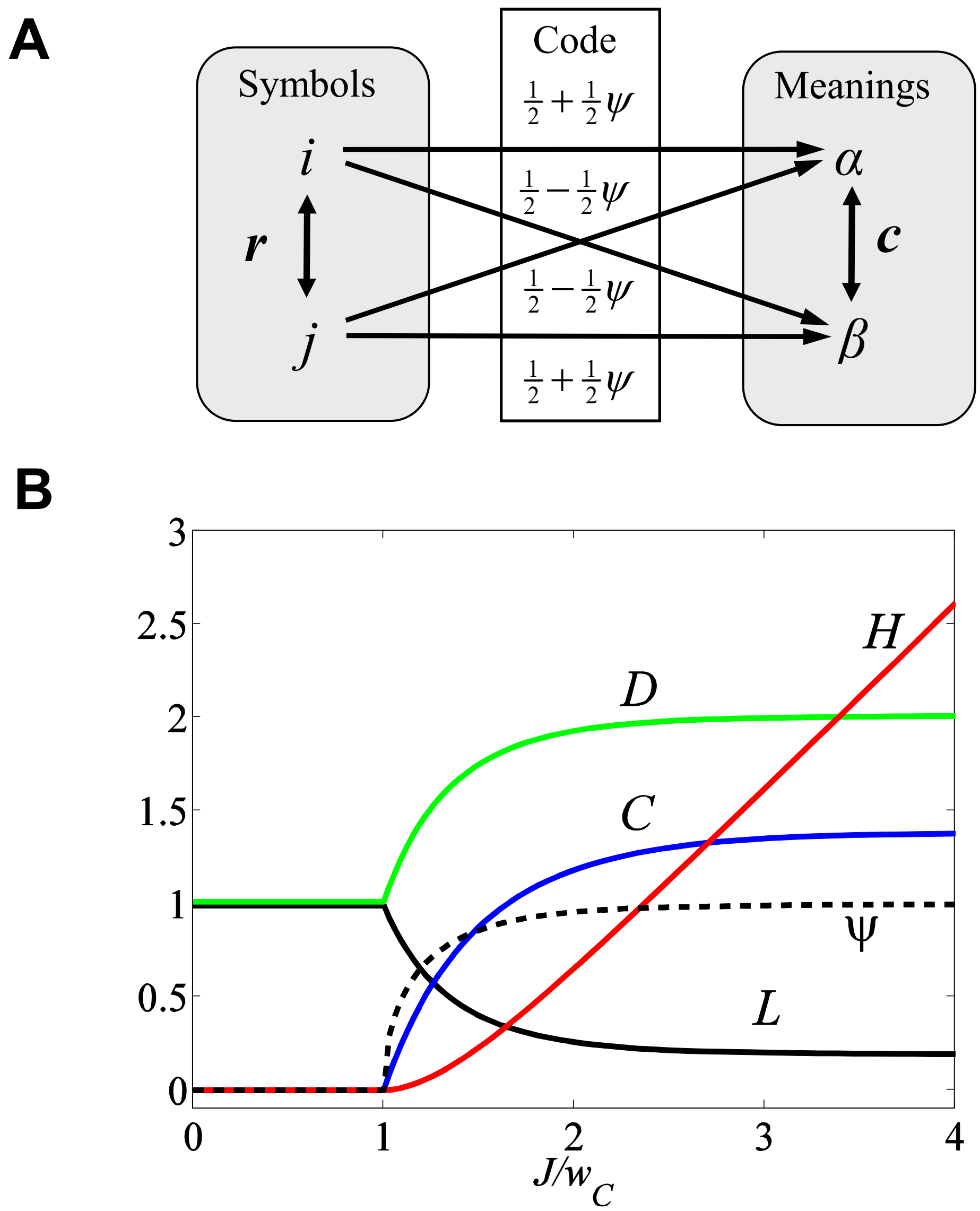

Fig. 2. **A 2-by-2 code**. **(A)** The code maps two symbols, *i* and *j*, to two meanings, $\alpha$ and $\beta$. The distance matrix ***c*** is specified by the parameter *c*, the average fitness load for replacing $\alpha$ by $\beta$ or $\beta$ by $\alpha$ (double-headed arrow, right), whereas the reading matrix ***r*** is specified by *r*, the average misreading probability (double-headed arrow, left). The arrows between symbols and meanings represent the code matrix ***p***, which is expressed as a function of the order parameter $\psi$. **(B)** The evolution of 2-by-2 code as a function of the ratio $J/w_C$: Plotted are the cost *C*, the diversity *D*, the error-load *L*, all scaled by *c*, the fitness *H*, scaled by $w_C$ (Eq. 8), and the order-parameter $\psi$ (Eq. 9). At high $w_C$ "temperature" (left) the system is in the non-coding state of uncorrelated symbols and meanings ($\psi = 0$). When $w_C$ decreases below a critical value $w_C{}^* = J$ (Eq. 10), the system undergoes a second-order coding transition. Following the coding transition the cost *C* increases, but this is compensated by the increasing diversity *D* and the decreasing error-load *L*, and thus the overall fitness *H* increases.

To demonstrate the coding transition, we examine a simple toy model of a coding system that maps two symbols, $i$ and $j$, to two meanings, $\alpha$ and $\beta$ (figure 2A). Due to the symmetry of the problem, the distance matrix $\boldsymbol{c}$, the reading matrix $\boldsymbol{r}$, and the code matrix $\boldsymbol{p}$, are all determined by one degree of freedom and can be written as

$$\text{(6)} \qquad \mathbf{c} = \begin{pmatrix} 0 & c \\ c & 0 \end{pmatrix}, \quad \mathbf{r} = \begin{pmatrix} 1-r & r \\ r & 1-r \end{pmatrix}, \quad \mathbf{p} = \tfrac{1}{2} \begin{pmatrix} 1+\psi & 1-\psi \\ 1-\psi & 1+\psi \end{pmatrix}.$$

The parameter $c$ measures the average fitness load of replacing $\alpha$ by $\beta$ or $\beta$ by $\alpha$ and the parameter $r$ is the average misreading probability, which is a measure for the "spin-spin" interaction range. The deviation of the code matrix from the uniform non-coding state $p_{i\alpha}$= 1/2 is measured by the order parameter $-1 \leq \psi \leq 1$. The error-load $L$, the diversity $D$, the cost $C$ and the resulting code fitness $H$ are obtained by substitution of Eq. 6 into Eqs. 1-4, which yields

$$\text{(7)} \qquad \begin{aligned} & L = c - c(1-2r)\psi^2, \quad D = c(1+\psi^2), \\ & C = (1+\psi)\ln(1+\psi) + (1-\psi)\ln(1-\psi), \\ & H = c(1-2r+w_D)\psi^2 - w_C\left[(1+\psi)\ln(1+\psi) + (1-\psi)\ln(1-\psi)\right], \end{aligned}$$

where an irrelevant constant was omitted from $H$. Evidently, the fitness $H$ is completely analogous, up to a minus sign, to the free energy of a mean-field Ising magnet (e.g. [13]). The first term in $H$ corresponds to the spin-spin interaction energy, which includes a ferromagnetic contribution due to the error-load and an anti-ferromagnetic interaction due to the diversity. The second term, which originates from the cost, corresponds to the entropy of the magnet. Within this analogy, the cost parameter $w_C$ is the temperature and the role of the spin-spin interaction strength $J$ is played by the parameter combination $J = c\ (1 - 2r + w_D)$. Thus, raising the interaction strength $J$ – by increasing $w_D$ or $c$, or by decreasing the error probability $r$ – is expected to increase the order and the correlation in the coding system.

Given the coding system parameters, $c$, $r$, $w_D$ and $w_C$, the optimal code is located at the maximum of the fitness. By calculating the extremum, $\partial H/\partial\psi = 0$, or directly by substitution of Eq. 6 into Eq. 5, we obtain a self-consistence relation for the optimal $\psi^*$, identical to that of the Ising magnet,

(8) $$\psi^* = \tanh\left(\frac{J}{w_C}\cdot\psi^*\right).$$

As in the Ising model, the code-matrix remains in the random non-coding state, $\psi^* = 0$, as long as the "temperature" $w_C$ is above a critical value $w_C^*$, which is equal to the interaction strength $J$,

(9) $$w_C^* = J = (1-2r+w_D)c.$$

A coding state, $\psi^* \neq 0$, emerges at a continuous, second-order coding transition at $w_C^*$ (figure 2B). Eqs. 8-9 indicate four possible pathways to approach the critical coding transition: (i) improving the reading accuracy (smaller $r$), for example by tailoring more specific binding sites (ii) raising the significance of diversity as measured by $w_D$, for example when the signals from the environment become more complex and require more bits for their representation (iii) increasing the fitness load $c$ of encoding a wrong meaning, and (iv) lowering the importance of cost, which is measured by $w_C$. The first three pathways are equivalent to strengthening the magnetic interaction while the fourth one is analogous to lowering the effective temperature.

Our discussion so far assumed implicitly that the evolution of the code follows the track of the optimum in the code fitness function. However, the coding system may get stuck at a metastable, sub-optimal state due to the ruggedness of the fitness landscape. In other words, there may exist, somewhere in the code fitness landscape, a superior, global optimum. Nevertheless, to reach this global optimum the coding system must cross deep 'valleys' or follow some intricate pathways. This may lead to ultraslow, 'glassy' dynamics. In this kind of almost frozen dynamics [1-3, 5, 15, 16], the distribution of local optima is much more important than the location of the global optimum. In addition, other effects of population dynamics, such as mutations and genetic drift, may drive the coding system towards suboptimal states [3].

### 4. A general criterion for the coding transition and the effect of symbol-space topology

The notion of a coding transition, which we demonstrated above in the simple 2-by-2 code, can be generalized by examining the stability of the non-coding state, $p_{i\alpha} = p_\alpha = 1/n_m$, with respect to small variations of the code-matrix $\delta p_{i\alpha} = p_{i\alpha} - p_\alpha$. This is equivalent to expanding the two-state

Ising magnet into an $n_m$-state Potts model. The order parameter in this case is variation $\boldsymbol{\delta p}$, whose $i$-th row corresponds to the deviation of average meanings encoded by the symbol $i$ from the uniform random state. This order-parameter describes the emergence of a code when $\boldsymbol{\delta p} \neq 0$. The fitness maximum at the non-coding state becomes unstable exactly at the point when the curvature tensor $Q_{i\alpha j\beta} = -(\partial^2 H/\partial p_{i\alpha}\partial p_{j\beta})$ ceases to be positive-definite. Taking the second derivatives of the fitness (Eq. 4) we find that the curvature at the non-coding state is $Q_{i\alpha j\beta}{}^* = 2(r_{ij} - w_D(1 - \delta_{ij}))c_{\alpha\beta} + n_m w_C\, \delta_{\alpha\beta}(\delta_{ij} - n_s^{-1})$. From an expansion of the variation $\boldsymbol{\delta p}$ in products of eigenvectors of $\boldsymbol{r}$ and $\boldsymbol{c}$, it is straightforward to see that the fitness becomes unstable at a critical temperature $w_C{}^*$ given by

$$w_C^* = 2n_m^{-1}\left(\lambda_r^* + w_D\right)\left|\lambda_c^*\right|, \tag{10}$$

where $\lambda_r{}^*$ is the second largest eigenvalue of $\boldsymbol{r}$ and $\lambda_c{}^*$, is the smallest eigenvalue of $\boldsymbol{c}$ ($\lambda_c{}^*$ is negative since the distance matrix is symmetric, real and traceless). For example, in the 2-by-2 code $\lambda_r{}^* = 1 - 2r$ and $\lambda_c{}^* = -c$, and substitution into Eq. 10 yields the critical $w_C{}^*$ given by Eq. 9. The general expression for the critical temperature (Eq. 10) indicates the same four pathways towards the coding transition demonstrated by the simple example: improving the reading accuracy (larger $\lambda_r{}^*$), raising the significance of diversity (larger $w_D$), lowering the cost of resources (smaller $w_C$) and increasing the load of replacing meanings (larger $|\lambda_c{}^*|$).

As mentioned above, the reading/misreading matrix $\boldsymbol{r}$ provides us with the geometric notion of a symbol-graph [2, 3, 10]. Within this geometry, the second-largest eigenvalue $\lambda_r{}^*$, which represents the emergent code, bears a special significance: The reading matrix $\boldsymbol{r}$ is related to the Laplacian of the symbol-graph $\Delta$ via $\Delta = I - \boldsymbol{r}$, where $I$ is the identity matrix [21]. It follows that $\lambda_r{}^*$ corresponds to the second-smallest eigenvalue $\lambda_\Delta{}^*$ of the Laplacian, $\lambda_r{}^* = 1 - \lambda_\Delta{}^*$ (We note that the simple 2-by-2 toy model considered above is degenerate in the sense that the second-smallest eigenvalue of the Laplacian, $\lambda_r{}^* = 1 - 2r$, is its only available excited-state). The Laplacian appears naturally in the noisy coding problem because it is the operator that describes random walk on the graph via misreading events that move the molecular reader along edges connecting confused symbols. The eigenvalue $\lambda_\Delta{}^*$ is the "mixing time", the slowest relaxation time in the system, which sets the time-scale for asymptotic decay to the uniform ground state. The corresponding eigenvector is known to be the *smoothest* of all excited modes of the graph, in accord with the intuitive physical notion that the modes with the lowest "energy" eigenvalues and frequencies are those of the largest wave-lengths.

To understand why it is the smoothest excited eigenmode that appears at the coding transition, it is useful to think of a physical analogy. In the coding system, the equivalent of energy is the error-load $L$. The mathematical expression for the error-load (Eq. 1) is somewhat similar to the elastic energy of a spring framework in the shape of the symbol graph. This may be further clarified if we consider the case of a Hamming distance $c_{\alpha\beta} = c(1 - \delta_{\alpha\beta})$, where the error load takes the form

$$L = \sum_{i,j} cr_{ij} \left( 1 - \sum_{\alpha} p_{i\alpha} p_{j\alpha} \right) = \sum_{i,j} cr_{ij} \left( 1 - \mathbf{p}_i \cdot \mathbf{p}_j \right).$$

The coefficients $cr_{ij}$ are the spring constants of the $(i, j)$ edges and the $i$-th row of the code matrix, $(p_{i\alpha}, p_{i\beta}, \ldots)$ is the $n_m$-dimensional position vector of the node $i$, $\mathbf{p}_i$. If the symbols connected by a "spring" encode different meanings then $\mathbf{p}_i \cdot \mathbf{p}_j = 0$ and the spring is stretched and costs error-load, whereas a "spring" connecting symbols that encode exactly the same meaning is loose and costs no error-load since $\mathbf{p}_i \cdot \mathbf{p}_j = 1$.

The present model predicts that the "softest" first-excited eigenvector $\boldsymbol{\delta p}^*$, the one that corresponds to the lowest non-zero eigenvalue of the Laplacian $\lambda_\Delta^*$, is the first mode to appear at the coding transition. It is a consequence of Courant's theorem that this eigenvector divides the graph into two connected regions, where in each of the region the eigenvector takes one sign [22-24]. In the non-negative region, $\boldsymbol{\delta p}^* \geq 0$, the symbols will tend to encode certain meanings, whereas in the other region the chance to encode these meanings will be less than the average, $\boldsymbol{\delta p}^* \leq 0$. Thus, the eigenvector $\boldsymbol{\delta p}^*$ represents a partition of the graph with minimal boundaries between regions of opposite tendency to encode certain meanings. Therefore, the emergent code is predicted to be smooth, in the sense that, on average, adjacent symbols encode similar meanings. This arrangement is advantageous because it minimizes the impact of misreading errors by decreasing the average difference between meanings encoded by adjacent symbols. For example, in a coding system of two meanings, say 'black' and 'white' it is clear that the error-load $L$ of an arrangement according to the lowest-excited mode $\boldsymbol{\delta p}^*$, where there are two large contiguous regions, one black and one white, is much smaller than the error-load of a chequered arrangement of the meanings.

The $\lambda_r^*$ smooth eigenmodes are known to be related to several structural and isoperimetric properties of the graph, such as the max-cut partition problem [21]. A relation between the eigenmodes of a membrane or a graph and its topology was considered in the classical work of Kac, 'Can one hear the shape of a drum?' [25]. Another relation in this spirit, between the multiplicity of the smooth eigenmodes and the topological problem of graph coloring has recently been discussed by us elsewhere [2, 10].

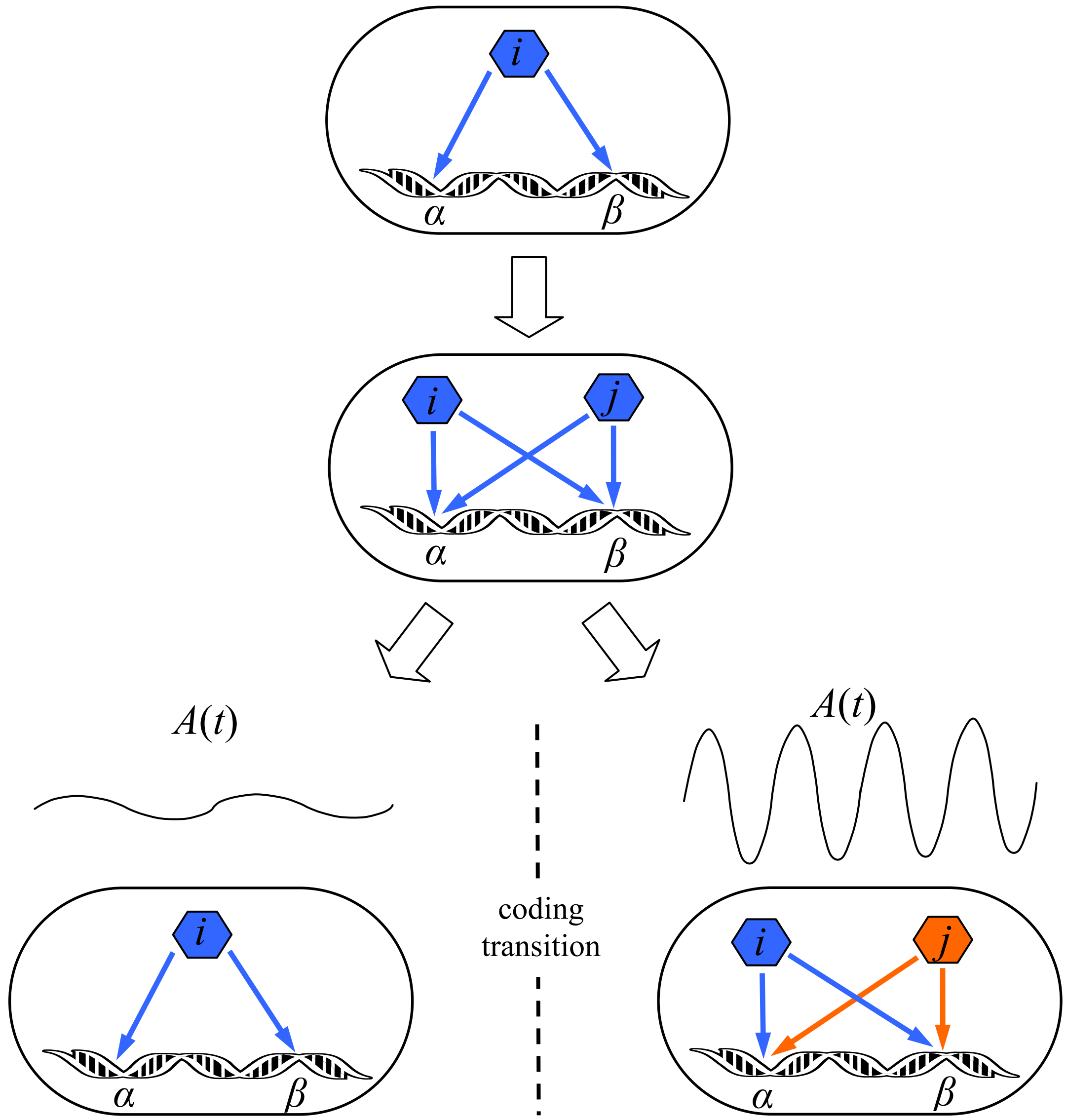

Fig. 3. **A sketch for an experimental measurement of the coding transition.** A transcription factor $i$ (blue hexagon) binds to two DNA binding sites $\alpha$ and $\beta$ (top). The experiment starts when the gene that controls the synthesis of $i$ is duplicated (middle). Then, one measures the evolutionary pathway of the bacteria in response to the gene duplication as a function of the environmental input $A(t)$ (bottom): It is expected that in a weakly fluctuating environment, the extra copy of the $i$ gene (denoted as $j$) is superfluous and it will eventually be deleted as the bacterium will return to the original single-copy state (bottom left). However, in a strongly fluctuating environment, the bacterium may differentiate the second copy of the gene (orange hexagon) to tailor a more efficient response of its regulatory network to the changing environment (bottom right). The benefit from preventing the synthesis of unnecessary proteins may surpass the cost of maintaining the extra gene and expressing the corresponding factor. The transition between the two possible fates is actually a coding transition (see text).

## 5. Discussion

The present model suggests a simple, intuitive mechanism for the emergence and evolution of noisy molecular codes. The quantitative scenario calls for experimental tests to falsify or verify its validity. As an example, let us consider the following sketch for an evolution experiment that involves the regulatory transcription network of bacteria (figure 3), disregarding at the moment the many potential obstacles that must be overcome in a real experiment. One may take a transcription factor $i$ that binds to two distinct DNA binding sites, $\alpha$ and $\beta$, and duplicate the gene that controls its synthesis. The aim of the experiment is to trace the evolutionary response of the bacteria to the duplication of the gene as a function of an environmental input $A(t)$, say the time-dependent concentration of a certain nutrient whose metabolism is controlled by $i$. It is expected that when the environment is constant or weakly fluctuating, an additional copy of the $i$ gene will provide the bacterium with hardly any benefit. Therefore, to rid of the cost involved with the maintenance of this redundant gene, denoted as $j$, it will eventually be deleted from the genome and the bacterium will return to the original single-copy state. However, when the environment is strongly fluctuating, the bacterium may utilize the second copy of the gene to tailor a more efficient response of its regulatory network to the changing environment. For example, by coevolving the two transcription factors and their binding sites the bacterium may control the expression levels such that certain proteins will be produced only when required. To accomplish this, the two transcription factors, $i$ and $j$, need to be differentiated via evolution to have different affinities and functions at the binding sites, $\alpha$ and $\beta$. The benefit from synthesizing proteins by demand may surpass the cost of maintaining the extra gene and expressing the corresponding transcription factor. At this point the organism will tend to retain the second gene.

The experimental scenario is similar in essence to our simple 2-by-2 toy model, and the transition from a single copy of the transcription factor to two copies that have different affinities to the binding sites is actually a coding transition. The experiment may be controlled by tuning the amplitude $|A|$ or the frequency $\omega$ of the environmental input $A$(t), which is analogous to varying the diversity parameter $w_D$ in the toy-model. Increasing $|A|$ is equivalent to increasing the significance of diversity, which drives the coding system towards the transition. Moreover, one may expect that $w_D$ will scale linearly with small variations of the experimental control parameter, $\delta w_D \sim \delta|A|$. Thus, one may imagine measuring the critical transition in response to varying the control parameter around its critical value, $\delta w_D \sim \delta|A| = |A| - |A|^*$, and quantitatively comparing the measurement to theoretical predictions (e.g. Eq. 7-8). In this regime, the fitness $H$ can be expanded to fourth-order, $H \sim (\delta w_D/w_C)\psi^2 - \psi^4$ and, as a result, the order-parameter scales like $\psi \sim (\delta w_D)^{1/2} \sim (\delta|A|)^{1/2}$. In other words, the model predicts for the order parameter the familiar square-root scaling of the mean-field Ising model. Of course, one needs to devise a method to accurately measure the order parameter, which is directly related to the affinities of the transcription factors to their DNA binding sites, and there might be many experimental difficulties and pitfalls. The simple model considered in the present work may also be useful to describe the evolution and emergence of other biological information systems that rely on noisy molecular recognition [26].